\documentclass{aa}  

\usepackage{graphicx}
\usepackage{txfonts}
\usepackage{color}
\usepackage{amsmath}    
\usepackage{lscape}
\usepackage{hyperref} 
\usepackage[switch]{lineno}
\let\oldequation\equation
\let\oldendequation\endequation
\renewenvironment{equation}
{\linenomathNonumbers\oldequation}
{\oldendequation\endlinenomath}
\usepackage{subfiles} 

\begin{document} 


   \title{Analysis of the possible detection of the pulsar wind nebulae of 
   PSR J1208-6238, J1341-6220, J1838-0537, and J1844-0346}
   \titlerunning{Analysis of the possible detection of the pulsar wind nebulae of four pulsars}
   
    \authorrunning{W. Zhang et al.}
   \author{Wei Zhang\inst{1,2},
          Diego F. Torres\inst{1,2,3},
          C. R. Garc\'{i}a\inst{1,2},
          J. Li\inst{4,5},
          Enrique Mestre\inst{1,2}
          }

   \institute{Institute of Space Sciences (ICE, CSIC), Campus UAB, Carrer de Magrans s/n, E-08193 Barcelona, Spain\\
              \email{zhang@ice.csic.es}
         \and
             Institut d’Estudis Espacials de Catalunya (IEEC), E-08034 Barcelona, Spain
             \and
             Institució Catalana de Recerca i Estudis Avanc˛ats (ICREA), E-08101 Barcelona, Spain
             \and
             CAS Key Laboratory for Research in Galaxies and Cosmology, Department of Astronomy, University of Science and Technology of China, Hefei, China
             \and
             School of Astronomy and Space Science, University of Science and Technology of China, Hefei, China
             }

   \date{Received November 27, 2023; accepted October 23, 2024}

  \abstract
   {Pulsar wind nebulae (PWNe) are a source of very high energy radiation that can reach up to tera-electron volts and even peta-electron volts. Our work uses the pulsar tree, a graph theory tool recently presented to analyze the pulsar population and select candidates of interest. }
   {We aim to discover detectable PWNe. We also aim to test to what extent the pulsar tree is able to group detectable PWNe despite only considering the intrinsic properties of pulsars.}
   {We selected four pulsars as tera-electron volt PWNe candidates based on their positions in the pulsar tree. Using observed and assumed ranges of values for relevant parameters, we anticipated the possible spectral energy distributions of the PWNe of four pulsars (PSR J1208-6238, J1341-6220, J1838-0537, and J1844-0346) via a detailed time-dependent leptonic model that was already found to be appropriate for describing almost all other detected nebulae.}
   {We estimated the likelihood of detection for the four candidates we studied by comparing the TeV fluxes predicted by the possible models with the sensitivities of different observatories. 
   In doing so, we provide context for analyzing the advantages and caveats of using the pulsar tree position as a marker for properties that go beyond the intrinsic features of pulsars that are considered in producing the pulsar tree.}
   {}


   \maketitle
%
\section{Introduction}
\label{sec1}

Pulsars are compact stars born in supernova explosions at the end of stellar evolution. They have fast rotation velocities and powerful magnetic and gravitational fields, which make them a laboratory for investigating extreme physical processes \citep{Gaensler_Slane06a, Slane:2017, Olmi_Bucciantini:2023}. Typically, more than 90\% of the energy that pulsars emit is released as a stellar wind formed by energetic particles. 

Most of the radiative models used to describe pulsar wind nebulae (PWNe) are applicable only to relatively young PWNe (where PWNe are not yet in contact with the reverse shock, so reverberation does not need to be considered in detail; see \cite{bandiera2023a, bandiera2023reverberation} and the references therein for discussion). According to existing models found to be in agreement with observations, the low-energy radiation of a PWN is mainly from synchrotron radiation generated by relativistic charges moving in a magnetic field, while the high-energy radiation is due to inverse Compton (IC) scattering from environmental photons -- mainly far infrared (FIR) photons (see, e.g., \cite{Atoyan1996,Aharonian1997,zhang2008, gelfand2009, dejager2009, tanaka2011, martin2012time, Bucciantini2011, martin2012time, torres2014time, vorster2013,torres2018,zhu2021}. Only Crab is known to be a self-synchrotron Compton (SSC) source. The SSC becomes relevant only for highly energetic (around 70\% of Crab) particle-dominated nebulae at low ages (of less than a few thousand years) located in a FIR background with a relatively low energy density \citep{Torres2013}. 

The model used in this work to describe the evolution of PWNe is based on TIDE, a one-zone leptonic time-dependent code capable of calculating both the energy spectrum and the dynamical evolution of young and middle-aged PWNe. Here, a simple thin-shell model is taken into account for the evolution of a PWN during the reverberation phase (see the incorporation of model details with time in \citealt{torres2014time,martin2016molecular,martin2022unique}). At the core of this model is seeking the solution of the equation \citep{ginzburg1964}
\begin{equation}
 \frac{\partial N(\gamma, t)}{\partial t}=-\frac{\partial}{\partial \gamma}[\dot{\gamma}(\gamma, t) N(\gamma, t)]-\frac{N(\gamma, t)}{\tau(\gamma, t)}+Q(\gamma, t) \text {. }
\end{equation}
The left side of this equation shows the evolution of the lepton distribution with time. The right side represents the variation of the particle energy with time due to the energy losses of different processes, including synchrotron, bremsstrahlung, IC, and adiabatic losses. Additionally, it accounts for the effect of particle escape on the lepton distribution, where $\tau(\gamma, t)$ is the escape time (assuming through Bohm diffusion) with $Q(\gamma, t)$ being the particle injection per unit energy per unit volume at a certain time $t$ and energy $\gamma$. 

The pulsar tree (see \cite{MST-I,MST-II} for full details) is a new way of visualizing the pulsar population, 
and it uses a graph theory tool known as a minimum spanning tree (MST) to represent the $P\dot P$ diagram. 
The MST is a graph that connects points in a multi-dimensional space. Each point (pulsar) is linked to at least another by an edge whose length is associated with a given distance. The latter is taken to be the Euclidean distance over the principal components of magnitudes representing the intrinsic characteristics of pulsars. The pulsar properties considered are the spin period ($P$), spin period derivative ($\dot{P}$), characteristic age ($\tau_c$), spin-down luminosity ($\dot{E}$), surface magnetic field ($B_s$), light cylinder magnetic field ($B_{lc}$), Goldreich-Julian charge density ($\eta_{GJ}$), and surface electric voltage ($\Phi_s$). The edges of an MST are chosen so that the sum of their lengths (equivalently, the sum of the total distance) when all nodes are linked and the whole population is connected is minimal. Graph theory shows that as long as distances are distinct, the MST is unique. Its definition intuitively implies that the MST is a classification technique. 

In this paper, we search for potential TeV PWNe based on their locations in the pulsar tree and use our one-zone PWN leptonic model to predict their possible energy spectrum characteristics. In Sect. \ref {method}, we introduce an analysis of MST where most known TeV PWNe are located in the lower-left region and choose candidates for further study, and subsequent sections show our results for the four objects chosen. Conclusions are given in Sect. \ref{conclusion}.

\section{Methodology}
\label{method}
\subsection{Global properties and candidate selection}
\label{MST}

\begin{figure*}
\centering
    \includegraphics[width=0.67\columnwidth]{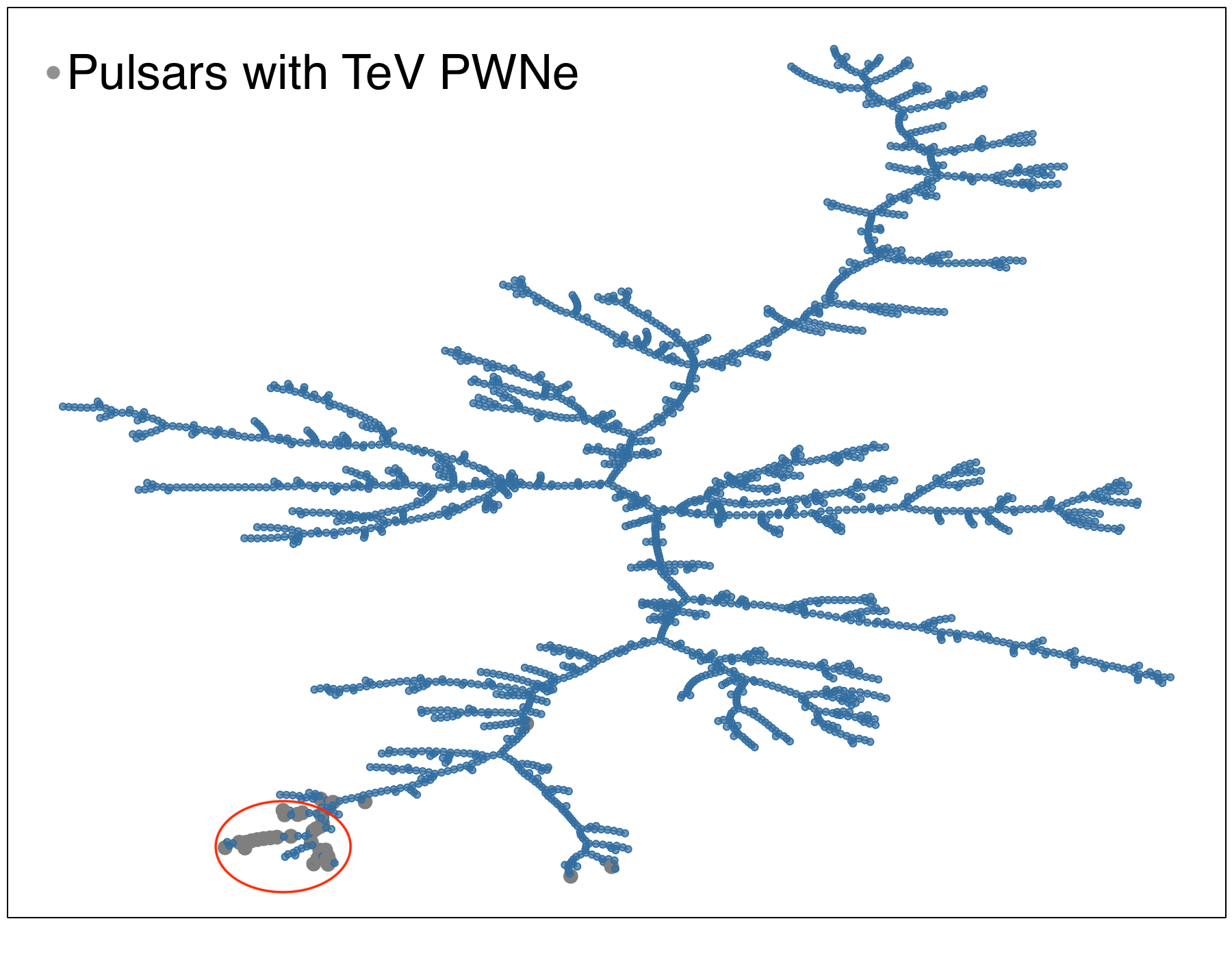}
    \includegraphics[width=0.67\columnwidth]{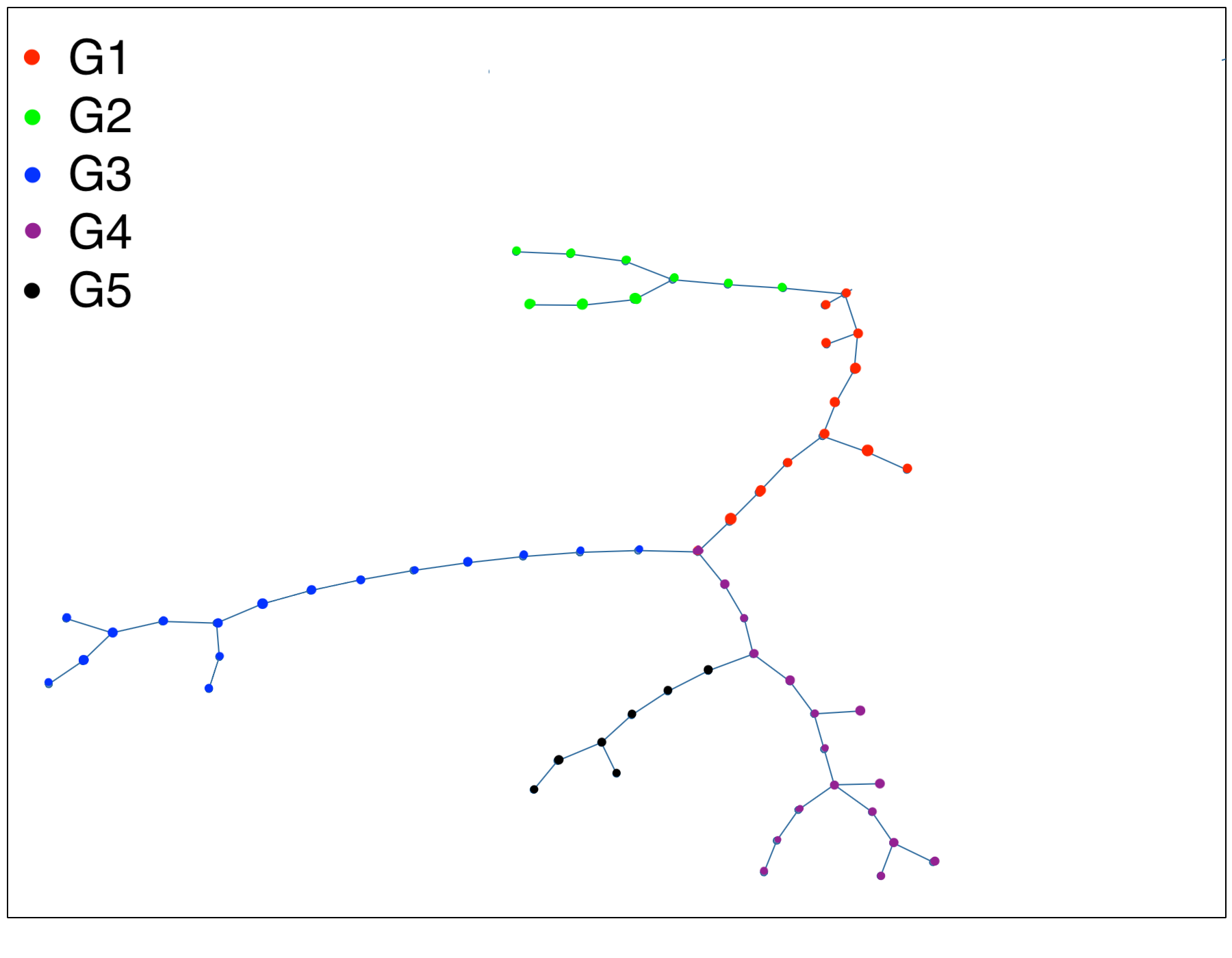}
    \includegraphics[width=0.67\columnwidth]{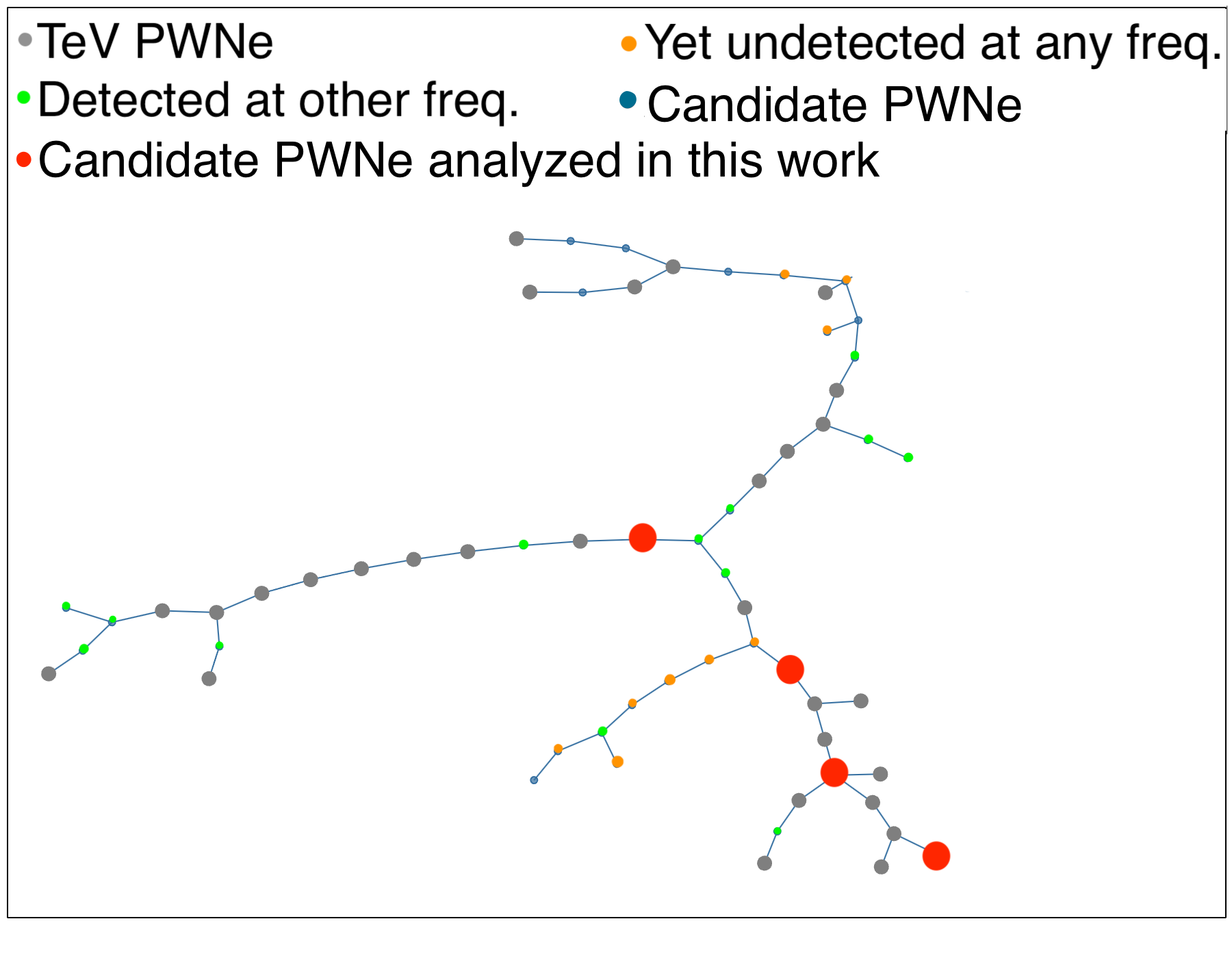}
       \caption{The pulsar tree and the positions of the selected pulsars in the pulsar tree. Left panel: Pulsars with detected TeV PWNe (gray dots) noted in the pulsar tree, the minimal spanning tree of the pulsar population.
Middle panel: Grouping of pulsars. The pulsars have been grouped in a way that allows us to comment on their different features (see text). This is a zoom-in of the bottom-leftmost part of the MST {(the area enclosed by the red ellipse)}, which is where most of the TeV PWNe are located. 
Right panel: Same zoom-in as in the middle panel, but we note the 29 pulsars with confirmed TeV PWNe, the 13 pulsars with confirmed PWNe at other frequencies, the nine pulsars without a PWN detected at any frequencies yet, and the ten remaining pulsars for which there are only unconfirmed candidate PWNe claimed. 
Four selected candidates analyzed in this work are noted in red, and these also belong to the group of ten pulsars. 
From top to bottom, the red points represent PSR J1844-0346, PSR J1341-6220, PSR J1838-0537, and PSR J1208-6238, respectively.}
    \label{fig1}
\end{figure*}

\begin{table*}
\setlength\tabcolsep{4pt}
        \centering
  \scriptsize
 \caption{Pulsars without confirmed TeV PWNe. }  

 \label{tabA1}
        \begin{tabular}{lcccccccc} 
                \hline
        \hline
                    PSR Name & $P$  &$\dot{P}$  &$\dot{E}$  &$\tau_c$  &PWN? &PWN / Composite  &Energy &Refs.\\
             & ($10^{-2}$ s) & ($10^{-14}$ s s$^{-1}$) & ($10^{36}$ erg s$^{-1}$) & (kyr) & &SNR Name &Band & (see also other refs. therein)\\
                        
  \hline
            J0540-6919 &5.06 &47.9 &146 &1.67 &Y &G279.7-31.5 &R, X, O, G$^?$ &\cite{manchester1993,mignani2012,bamba2022spectral}\\
            J0940-5428 &8.75 &3.29 &1.93 &42.2 &N & & &\\
            J1015-5719 &14.0 &5.74 &0.828 &38.6 &Y &G283.1-0.59 &R &\cite{ng2017discovery}\\
            J1044-5737 &13.9 &5.46 &0.803 &40.3 &N & & &\\
            J1048-5832 &12.4 &9.61 &2.00 &20.4 &Y & &X &\cite{gonzalez2006}\\
            J1111-6039 &10.7 &19.5 &6.35 &8.66 &Y &G291.0-0.1 &R, X, G &\cite{slane2012broadband}\\
            J1112-6103 &6.50 &3.15 &4.53 &32.7 &$?$ & &X$^?$, G$^?$ &\cite{townsley2011integrated,ackermann2016}\\
            J1124-5916 &13.5 &75.3 &11.9 &2.85 &Y &G292.04+1.75 &R, X, G$^?$ &\cite{ajello2017,park2007,gaensler2003}\\
            J1135-6055 &11.5 &7.93 &2.06 &23.0 &Y &G293.8+0.6 &X, R$^?$ &\cite{zhang2019chandra,bordas2020}\\
            J1203-6242 &10.1 &4.41 &1.71 &36.1 &N & & &\\
            J1208-6238 &44.1 &327 &1.51 &2.14 &? & &X$^?$ &\cite{bamba2020low}\\
            J1341-6220 &19.3 &25.3 &1.38 &12.1 &? &G308.8-0.1 &R$^?$, G$^?$ &\cite{acero2015fermi,green1997continuation}\\  
            J1400-6325 &3.12 &3.89 &50.7 &12.7 &Y & &X, R &\cite{bamba2022spectral,renaud2010}\\
            J1410-6132 &5.01 &3.20 &10.1 &24.8 &? &G312.4-0.4 &X$^?$ &\cite{doherty2003}\\
            J1524-5625 &7.82 &3.90 &3.21 &31.8 &N & & &\\
            J1637-4642 &15.4 &5.93 &0.640 &41.2 &? &HESS J1640-465 &X$^?$,G$^?$, T$^?$ &\cite{slane2010fermi,lemiere2009,aharonian2006}\\
            J1730-3350 &14.0 &8.48 &1.24 &26.1 &N & & &\\
            J1747-2958 &9.88 &6.13 &2.51 &25.5 &Y &Mouse &R,X &\cite{klingler2018}\\
            J1801-2451 &12.5 &12.8 &2.59 &15.5 &Y &Duck &R,X &\cite{kaspi2001,gaensler2000}\\
            J1813-1246 &4.81 &1.76 &6.24 &43.4 &? &HESS J1813-126 &X$^?$, T$^?$ &\cite{marelli2014puzzling,tibolla2022pulsar}\\
            J1826-1256 &11.0 &12.1 &3.58 &14.4 &Y &Eel &X, G$^?$, T$^?$ &\cite{burgess2022eel,roberts2007,breuhaus2022}\\
            J1837-0604 &9.63 &4.52 &2.00 &33.8 &? &LHAASO J1839-0545 &T$^?$ &\cite{de2022potential,cao2021}\\
            J1838-0537 &14.6 &47.2 &6.02 &4.89 &? &HESS J1841-055 &T$^?$ &\cite{magic2020studying,aharonian2008}\\
            J1844-0346 &11.3 &15.5 &4.25 &11.6 &? &HESS J1843-033 &T$^?$ &\cite{sudoh2021highest,devin2021multiwavelength,amenomori2022measurement}\\
            J1932+2220 &14.5 &5.76 &0.754 &39.8 &N & & &\\
            J1934+2352 &17.8 &13.1 &0.908 &21.6 &N & & &\\
            J1935+2025 &8.01 &6.08 &4.66 &20.9 &? &G054.1+00.3 &X$^?$, T$^?$ &\cite{temim2010,xia2023}\\
            J2021+3651 &10.4 &9.57 &3.38 &17.2 &Y &G75.2+0.1 &R, X, T$^?$ &\cite{roberts2008,woo2023,abeysekara2020multiple}\\
            J2022+3842 &4.86 &8.61 &29.6 &8.94 &Y &G076.9+1.0 &R, X &\cite{arzoumanian2011,marthi2011}\\
            J2111+4606 &15.8 &14.3 &1.44 &17.5 &N & & &\\
            J2229+6114 &5.16 &7.83 &22.5 &10.5 &Y &Boomerang &R, X, G, T$^?$ &\cite{magic2023,pope2024}\\
            J2238+5903 &16.3 &9.70 &0.889 &26.6 &N & & &\\
                \hline
        \end{tabular}
\tablefoot{The parameters in columns 2, 3, 4, and 5 are from the ATNF catalog\footnote{\url{http://www.atnf.csiro.au/research/pulsar/psrcat}} \citep{manchester2005australia}. The "Y" in column 6 indicates that its PWN has been confirmed in one or several bands; an "N" means that the PWN hasn't been observed yet to our knowledge; and a question mark means that there is a possible PWN counterpart in some band or bands but that it has not yet been confirmed. In the eighth column (PWN energy band), "R, O, X, G, T" stand for radio, X-rays, visible, GeV, and TeV energy ranges. }
\end{table*}

The first panel of Fig. \ref{fig1} shows the pulsar tree \citep{MST-I}. Online access to this figure (and tools to zoom and label each node) is also available.\footnote{\url{http://www.pulsartree.ice.csic.es/pulsartree}} As can be seen in the first panel of Fig. \ref{fig1}, most of the pulsars with detected TeV PWNe \citep{hess2018population} are clustered in the lower-left region of the pulsar tree. There are 61 pulsars in this region, shown in the second panel of Fig. \ref{fig1}. Twenty-nine of them have confirmed TeV PWNe, 13 pulsars have confirmed PWNe detected in other bands, and another ten pulsars have been noted to have possible PWN counterparts at some frequencies that have not yet been confirmed. We did not find any reports regarding possible PWNe in any bands for the remaining nine pulsars. The overdensity of TeV-detected PWNe in a small part of the MST is quite notorious. If TeV PWNe were placed randomly in the MST, the probability that this overdensity existed would be negligible. The MST is joining similarly energetic and relatively young sources that are prone to have TeV PWN behavior. However, we recall that no connection with the environment or with the progenitor is part of the distances underlying the MST.

Table \ref{tabA1} lists the (61-29=) 32 pulsars without confirmed TeV PWNe. We used the visually obvious branches appearing in the pulsar tree in this region to divide the pulsars into five groups for further study (see the middle panel of Fig. \ref{fig1}). The sources located at the junctions of these branches are included in the group to which they are the most similar. For example, PSR J1826-1256, which is at the junction of G1, G3, and G4, has six (five) [three] of the eight magnitudes considered that are larger or smaller than the corresponding values of the sources in G1 (G3) [G4]. So we include it in G4. None of the sources in G5 have a detected TeV PWNe. Also, we observed that the values of $\dot{E}$ and $\Phi_s$ (proportional to $\dot P/P^3$) in G5 are smaller than those in any other group. Of the six pulsars in G1 with the smallest values in $\dot{E}$ and $\Phi_s$, only PSR J1809-1917 has a detected TeV PWN. Similarly, the G4 pulsars
PSR J1341-6220, PSR J2111+4606, and PSR J1208-6238, which have the smallest $\dot{E}$ and $\Phi_s$ (PSR J2111+4606 does not have a PWN detected in any band so far), and the other two sources have only a possible PWN in X-ray and radio, respectively. The tree seems to group together pulsars that are less likely to have detectable PWNe. 

From the sources that do not have a confirmed TeV PWN already, we selected candidates for further study from possibly young sources with $\tau_c<15$ kyr, which still lack a detailed radiative analysis from which predictions can already be derived. We chose PSR J1838-0537, PSR J1208-6238, PSR J1844-0346, and PSR J1341-6220 under these conditions. The positions in the pulsar tree of these four sources are also shown in Fig. \ref{fig1}. Our subsequent investigation sought to determine whether these four pulsars could be producing plausibly observable PWNe, or if models similar to those used to describe other well-known PWNe predict that they should be undetectable.

\subsection{Prior TeV observations and PWN models}
\label{analyse}

The positions of all four candidates are covered in the H.E.S.S. Galactic Plane Survey (HGPS; \citealt{abdalla2018hess}), and no PWNe was claimed at its sensitivity for the regions of PSR J1208-6238 and PSR J1341-6220. HESS J1841-055 and HESS J1843-033 are reported near PSR J1838-0537 and PSR J1844-0346, respectively, and both of them are regarded as one of a few plausible counterparts, based on being spatially coincident with the very high energy (VHE) source, albeit without certainty. Thus, we note the sensitivity of the HGPS at the position of the pulsars as an upper limit in their corresponding spectral energy distributions (SEDs) below, when appropriate. 

We simulated the possible SEDs of the putative PWNe of all four candidates using our one-zone leptonic model. The relevant physical parameters involved are shown in Table \ref{tab2}. The true age, $t_{age}$, of the pulsar was assumed to range from either 0.7 to 1.3 times $\tau_c$ or as wide a range as possible in order to secure a positive initial spin-down age $\tau_0$ given the adopted braking index. The base energy densities of soft photon fields, FIR and near infrared (NIR) photons, were adopted from the GALPROP code according to the coordinates and distances of these sources \citep{porter2022galprop}. However, considering that the GALPROP model estimates could underpredict the densities at the scale relevant for PWNe (see the discussion in \cite{torres2014time} and references therein), the energy densities were assumed to range from one to three times these values. 

We constructed models that range over nine parameters, as described in Table \ref{tab2}. We combined the minimum and maximum values of each of the nine parameters, which gave us 2$^{9}$=512 models, and then we added a group of 1000 model realizations by taking random values within all of these intervals, yielding a total of 1512 possible models for each source. These models span the range of possible SEDs given the uncertainties, and we used them to assess how likely it is to detect these PWNe in case they behave in a manner similar to the ones that are already known. We note that we considered a wide span of ejecta masses.
A lower $M_{ej}$ would result from a zero-age main sequence eight solar mass progenitor and could also be possible for windy progenitors such as luminous blue variables and Wolf-Rayet stars but would not be the usual case. Variations in the explosion energy are expected to be mild for such young pulsars, where approximately, for instance, $R_{pwn}(t) \propto \sqrt[5]{L_0t/E_{sn}}V_{ej}t$, and $V_{ej}=\sqrt{10E_{sn}/(3M_{ej})}$ while in free expansion.

\section{Results}
\label{results}

\subsection{PSR J1208-6238}
\label{secJ1208}

A very young, energetic, and highly magnetized radio-quiet $\gamma$-ray pulsar, PSR J1208-6238 has a spin-down luminosity of $1.51 \times 10^{36}$ erg s$^{-1}$. Its spin frequency and its first-order derivative are 2.2697 Hz and $-16.8427\times10^{-12}$ Hz/s, respectively, resulting in a characteristic age $\tau_c$ of 2672 years, and its braking index is $n = 2.598$, based on five years of Fermi-LAT observations \citep{clark2016braking}. The inferred dipolar magnetic field is relatively high, $3.8 \times 10^{13}$ G. The distance ($D$) to this pulsar is unknown, and we adopted 3, 6, and 10 kpc as different assumed values. Since the discovery of the pulsar, no evidence of a PWN has been observed at any wavelength until \cite{bamba2020low} reported X-ray emission from a possible PWN. The confidence level of this detection is 4.4 $\sigma$, but the association with a PWN is unclear (see the discussion of these authors). Their results would imply a conversion factor of $\dot{E}$ to the X-ray luminosity, which is unusually small ($< 10^{-4}\, D_3^2$, where $D_3$ represents the distance of the pulsar in units of 3 kpc). 
%

\begin{table*}
        \centering
  \scriptsize
        \caption{Physical magnitudes. } 
        \label{tab2}
        \begin{tabular}{llllll} 
                \hline
        \hline
                Parameters & Symbol & J1208-6238 &J1341-6220 &J1838-0537 &J1844-0346 \\
                \hline
         Measured, assumed, or derived parameters: \\
                        \hline
                
                Period (s) & $P$ &0.441 &0.193 &0.146 &0.113 \\
        Period derivative $(s~s^{-1})$ &$\dot{P}$ &$3.27\times10^{-12}$ &$2.53\times10^{-13}$ &$4.72\times10^{-13}$ &$1.55\times10^{-13}$\\
                Characteristic age (kyr) &$\tau_c$ &2.14 &12.1 &4.89 &11.6\\
                Spin-down luminosity now (erg s$^{-1})$ &$\dot{E}$ &$1.51\times10^{36}$ &$1.38\times10^{36}$ &$6.02\times10^{36}$ &$4.25\times10^{36}$ \\
                Distance (kpc) & $d$ &3, 6, 10  &12.6 &2.0  &2.4, 4.3, 10 \\
        Initial spin-down age &$\tau_0$ &$({2\tau_c})/({n-1}) - t_{age}$ & & &\\
        Initial spin-down luminosity (erg s$^{-1})$ &$L_0$ &$\dot{E}/(1+{t_{age}}/{\tau_0})^{-(n+1)/(n-1)}$ & & &\\
   \hline               
        Parameter ranges: \\
                        \hline
        Age ($\tau_c$) &$t_{age}$ &[0.7, 1.2] &[0.7, 1.3] &- &- \\
        ISM density (cm$^{-3}$) &$n_{ism}$ &[0.1, 1] &- &- &- \\
        Break energy &$\gamma_b$ &[$10^5, 10^7$] &- &- &-\\
        Low energy index &$\alpha_1$ &[1.0, 1.6] &- &- &-\\
                High energy index &$\alpha_2$ &[2.2, 2.8] &- &- &-\\
                Ejected mass $(M_{\odot})$  &$M_{ej}$ & {\bf [5, 15]} &- &- &-  \\
                Magnetic energy fraction  &$\eta$ &[0.02, 0.04] &- &- &-\\
                FIR energy density (eV cm$^{-3}$) &$U_{fir}$ &[0.55, 1.65] &[0.29, 0.87] &[0.82, 2.46] &[0.82, 2.46]\\
                NIR energy density (eV cm$^{-3}$) &$U_{nir}$ &[0.82, 2.46] &[0.51, 1.53] &[1.06, 3.18] &[1.06, 3.18] \\
        Breaking index & $n$ &2.598 (measured) &2/2.5/3 &- &- \\
        Minimum energy at injection &$\gamma_{min}$ &1 &- &- &- \\
        Containment factor  &$\epsilon$ &0.5 &- &- &- \\
        SN explosion energy (erg) &$E_{sn}$ &$10^{51}$ &- &- &- \\
        CMB temperature (K)  &$T_{cmb}$ &2.73 &- &- &-\\
                CMB energy density (eV cm$^{-3}$)  &$U_{cmb}$ &0.25 &- &- &-\\
        FIR temperature (K)  &$T_{fir}$ &70 &- &- &-\\
                NIR temperature (K)  &$T_{nir}$ &3000 &- &- &-\\
                \hline
        \end{tabular}
 \tablefoot{The symbol "-" in the table means the value or range is the same as noted in the previous column. To make sure the $\tau_0$ will not be negative, the braking index for the sources except J1208-6238 was assumed to be 2/2.5/3, which means that with the increasing of the assumed value for $t_{age}$ from 0.7 to three times $\tau_c$, the braking index would be assumed as 3, 2.5, and 2, respectively.}
\end{table*}
\begin{figure*}
    \includegraphics[width=.67\columnwidth]{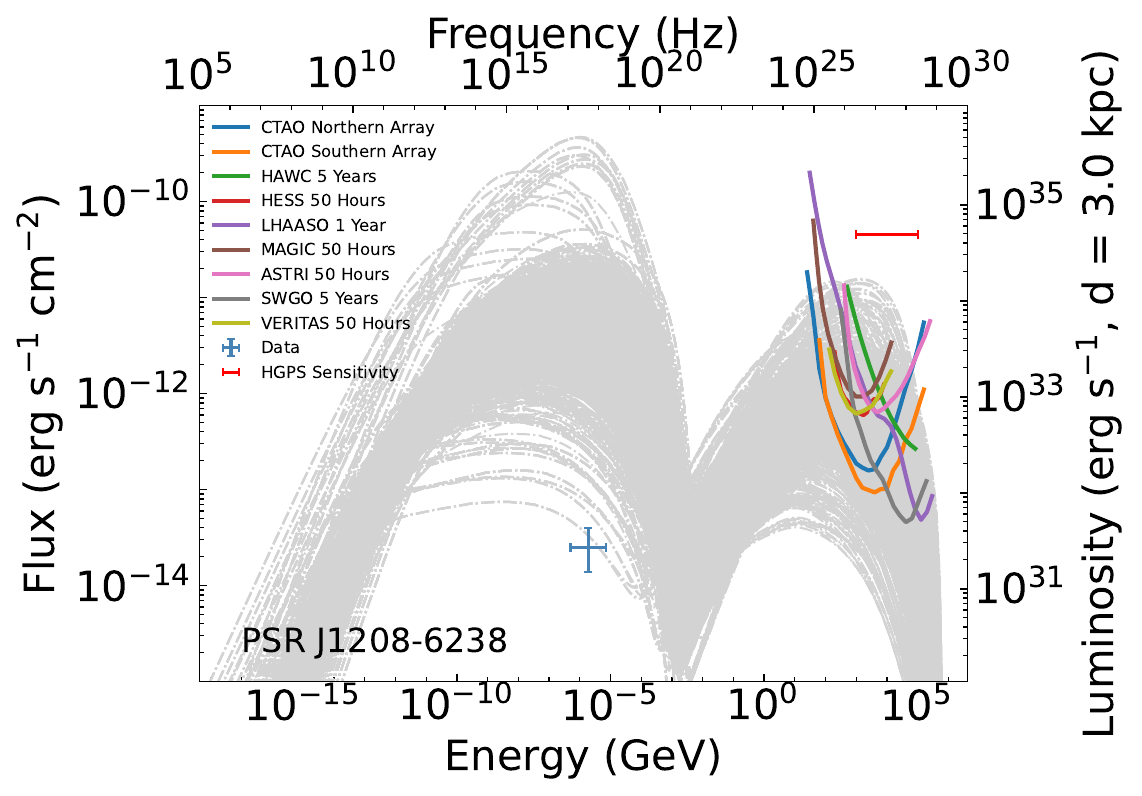}
    \includegraphics[width=.67\columnwidth]{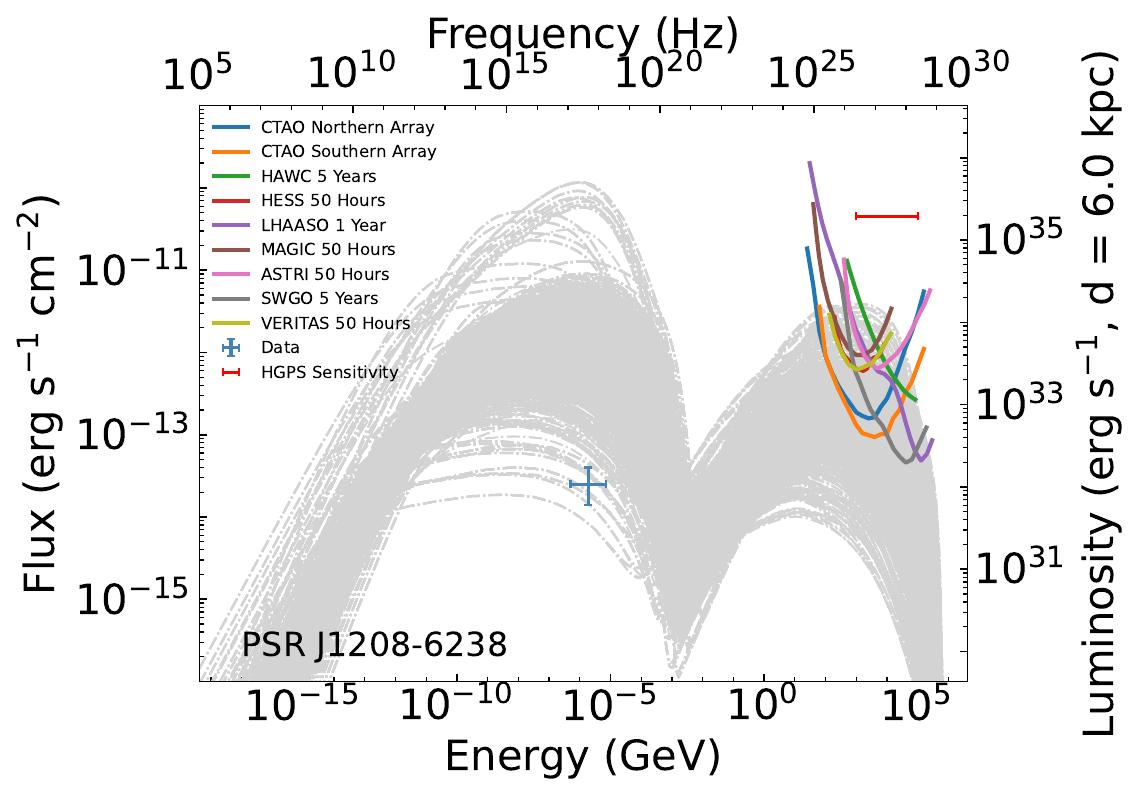}
    \includegraphics[width=.67\columnwidth]{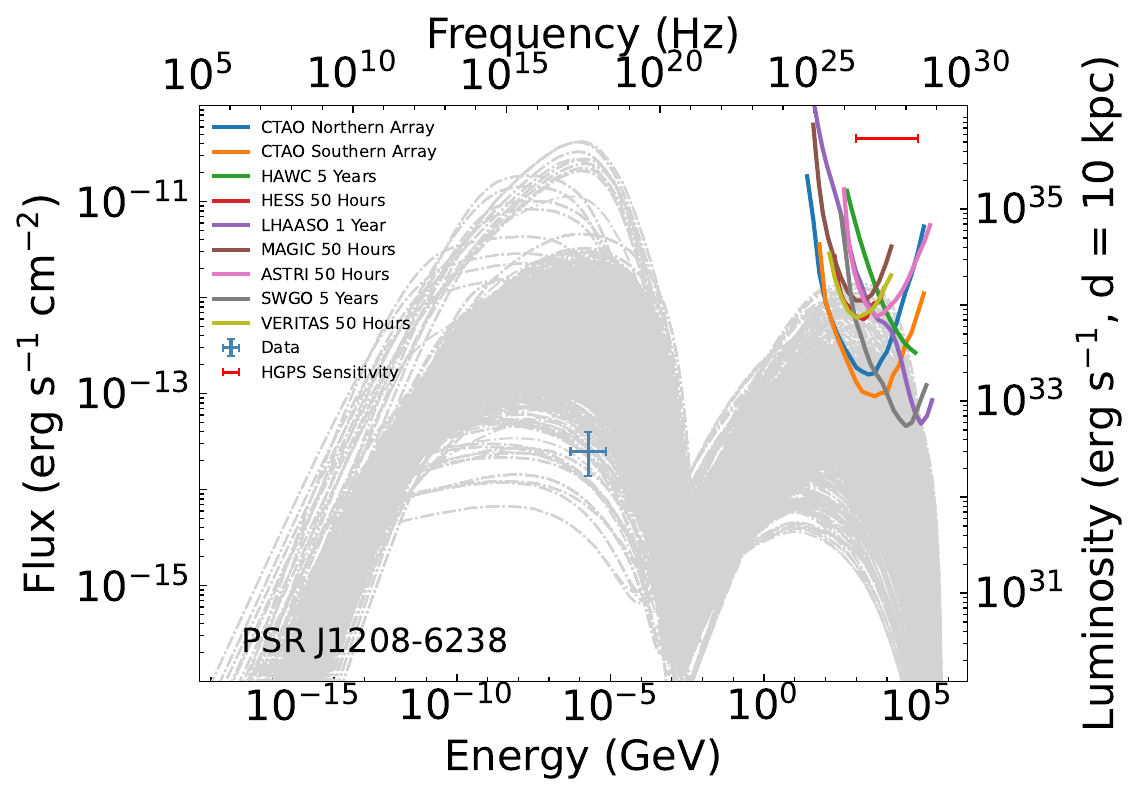}
    \includegraphics[width=.67\columnwidth]{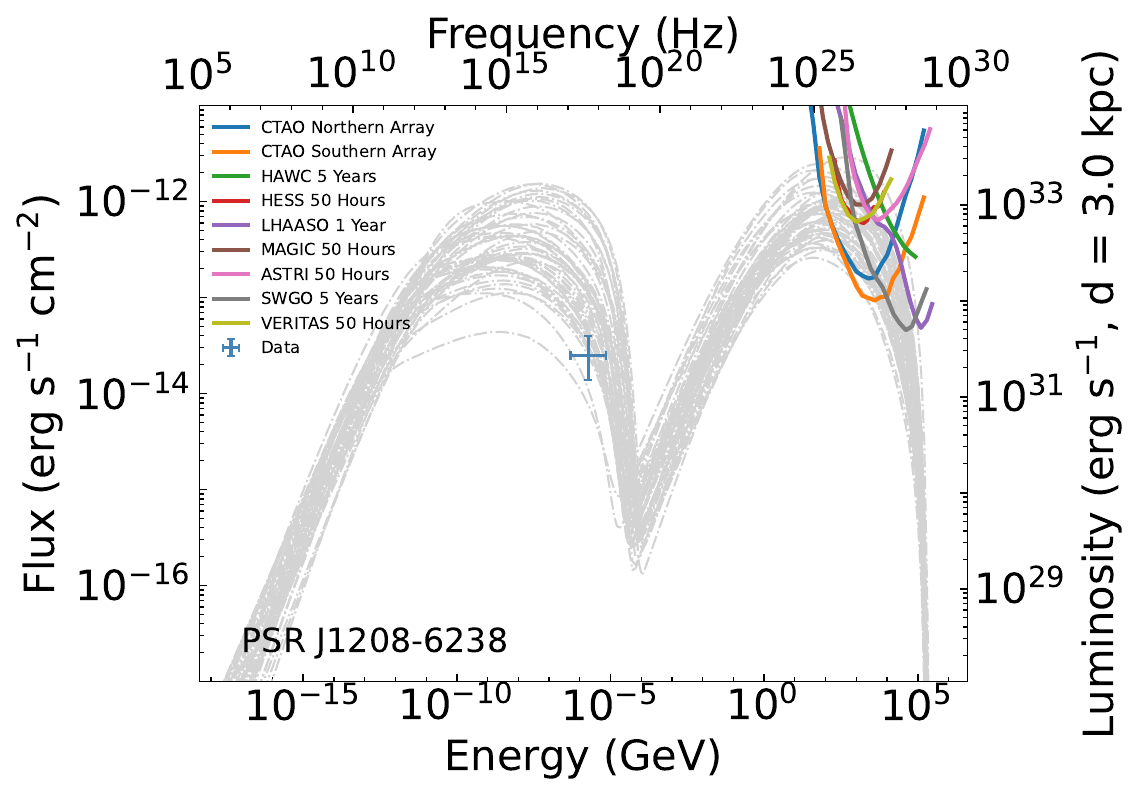}
    \includegraphics[width=.67\columnwidth]{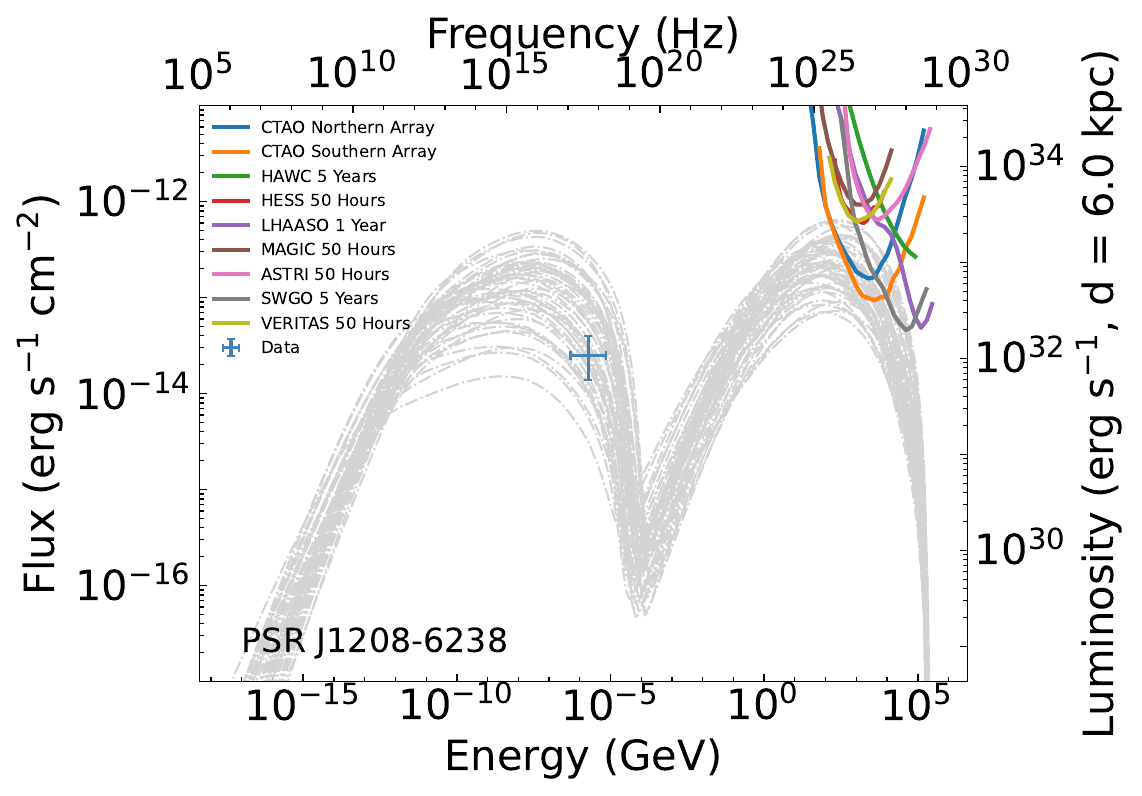}
    \includegraphics[width=.67\columnwidth]{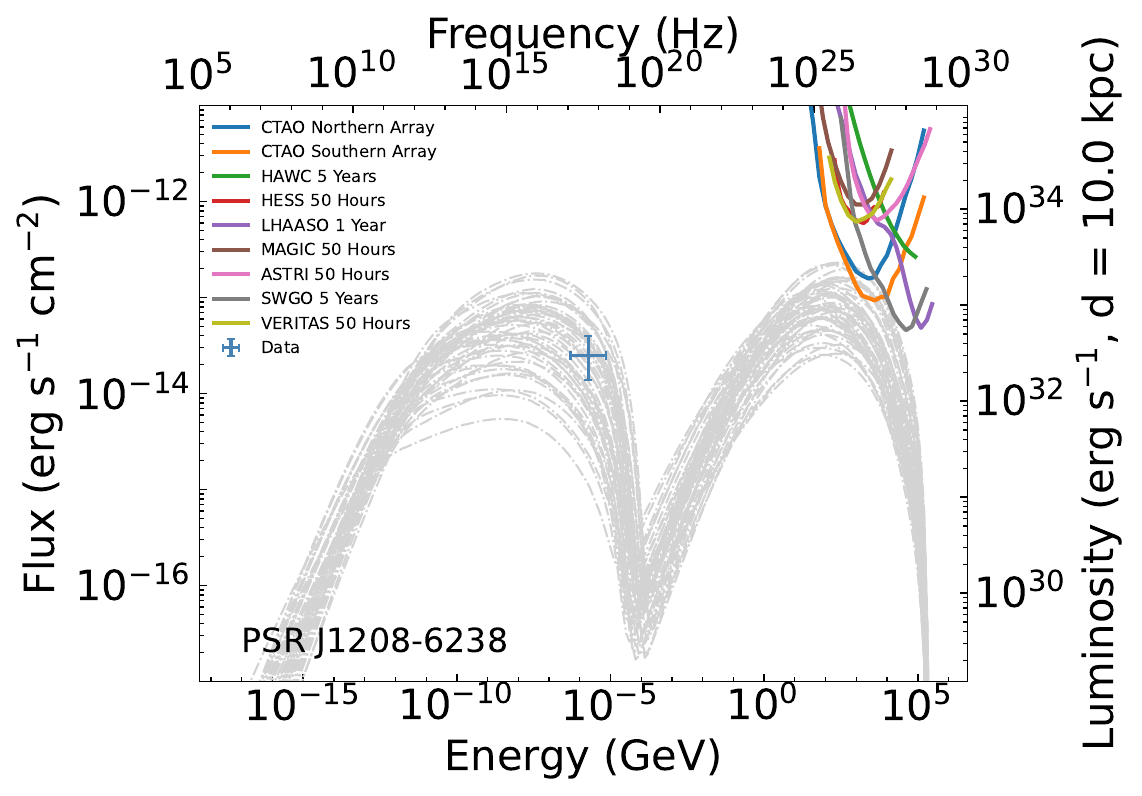}
   \caption{
Predicted set of SEDs for the PWNe of PSR J1208-6238. First three panels: Predicted SEDs for the PWN of PSR J1208-6238 with distances assumed to be 3 kpc, 6 kpc, and 10 kpc, respectively. 
The sensitivities of several instruments are from the CTA public website\protect\footnotemark. 
Last three panels: Fifty predicted SEDs for the PWN of PSR J1208-6238. These predictions were obtained by {fixing $\eta$ to 0.003 (for 3 kpc) or 0.004 (for 6 and 10 kpc)} and randomly selecting the other eight parameters in the ranges shown in Table \ref{tab2}. 
}
    \label{sed_J1208}
\end{figure*}
\footnotetext{\url{https://www.ctao.org/for-scientists/performance/}}

Figure \ref{sed_J1208} shows the model realizations under different distance assumptions. At 3 kpc (this distance is also adopted in \citealt{clark2016braking} and \citealt{bamba2020low}), most of the model realizations would lead to a detectable PWN surrounding PSR J1208-6238, despite all of the models concluding that it would be undetectable in the HGPS. However, the realizations would all conflict with the X-ray observations of the region unless the size of the PWN is different from the X-ray region that is covered. Indeed, in the analysis of \cite{bamba2020low}, the authors used a ring centered on the pulsar as the possible PWN region, which had an inner radius of $\sim 5^{\prime\prime}$ and an outer radius of $\sim 9^{\prime\prime}$ (i.e., 0.13 pc outer radius at an assumed distance of 3 kpc). However, in our models, $R_{pwn}$ ranges from 0.3 to 4 pc at the same distance. If the X-ray size of this PWN is underestimated, the X-ray flux may also be underrated. 

In the case of 6 kpc, only eight models do not exceed the X-ray upper limit, but all of them would be below the sensitivity of any available TeV observational equipment, rendering the PWN undetectable. Even at 10 kpc (which is still within the maximum distance of 18.9 kpc reported in \cite{clark2016braking} by assuming that this pulsar is at the Galaxy edge for the given line of sight), this number rises to 41, representing 2.71\% of the total 1512 models, all of which are also undetectable in the TeV band. This result would be consistent with the fact that no TeV sources have been observed in this region so far. If this PWN behaved similarly to others in the known sample, we would likely find no TeV counterpart from it.

If the X-ray flux is indeed to be taken as an upper limit of the PWN, we also considered whether such a low flux could be just the result of a more extreme value of the magnetic fraction. Such low values were also needed in the modeling of some other PWNe (see the compilation provided by  \cite{Abdelmaguid2023}). Here, the value of $\eta$ needs to be less than 0.004 for 3 kpc and less than 0.005 for 6 and 10 kpc (as shown in the bottom panels of Fig. \ref{sed_J1208}) to respect the X-ray upper limit and make the TeV fluxes reach the sensitivity of the southern array of the Cherenkov Telescope Array (CTA) (which we refer to as S-CTA). These values of $\eta$ are much lower than the typical value found in other nebulae, which is typically around a few percent. This does not seem to offer a solution to the fact that it seems unlikely that PSR J1208-6238 would produce a detectable TeV PWN. Only the S-CTA, or H.E.S.S., can make a dedicated observation of this source, which we nevertheless promote as a way of testing these conclusions. 

Finally, if we consider the PWN to be a diffuse source of uncertain size and the X-ray data for a smaller PWN in \cite{bamba2020low} to provide an upper limit on its surface brightness, our models produce a surface brightness consistent with the X-ray constraint in about 45\% of all 1512 trials, and most of them are visible with S-CTA. However, we note the caveat that this assumes that the source is extended, dim, and uniformly diffuse, which indeed promotes fewer constraints onto the models and does not appear to be the case in X-ray observations of younger PWNe, where the X-ray emission is rather peaked.

\subsection{PSR J1341-6220}  
\label{secJ1341}

PSR J1341-6230 is a Vela-like radio pulsar discovered by the Parkes radio telescope with a spin period of 0.19 s, a characteristic age of $\tau_c = 12.1$ kyr, and a spin-down luminosity of $\dot{E} = 1.38 \times 10^{36}$ erg s$^{-1}$ \citep{manchester1985search,manchester2005australia}. Its distance is taken as 12.6 kpc from the ATNF catalog, based on the YMW16 electron density model. The weak X-ray counterpart of this pulsar was also detected in X-rays \citep{kuiper2015soft}. Frequent glitch phenomena were detected for this source; for instance, \cite{lower2021impact} reported 15 glitches. 

\cite{wilson1986x} gave an upper limit for the X-ray flux of PSR J1341-6220 on the order of $2.5 \times 10^{-13}$ erg s$^{-1}$ cm$^{-2}$ ($0.2-3.5$ keV). This pulsar is also listed as a radio pulsar with an X-ray counterpart in table 3 of \cite{Kaplan2004} with an about one order of magnitude smaller flux than quoted by \cite{wilson1986x}; however, it is unclear where this limit comes from, as they provide no analysis and quote an unpublished reference. \cite{Prinz2015} reported that the pulsar was in the field of view of the XMM-Newton telescope for about 42 ksec during two different epochs with different off-axis angles. Their analysis revealed two close sources (barely distinguishable and just above 2 keV; see their Fig. 1). They provided an upper limit for the harder of these two sources only. Although it is clear that there is no bright X-ray PWN around this pulsar, assigning a specific upper limit in order to discard models with it is, at this time, risky.
We promote the undertaking of a dedicated X-ray observation of this pulsar with Chandra to finally determine its nebular characteristics, if any.

\begin{figure}
\centering
    \includegraphics[width=.99\columnwidth]{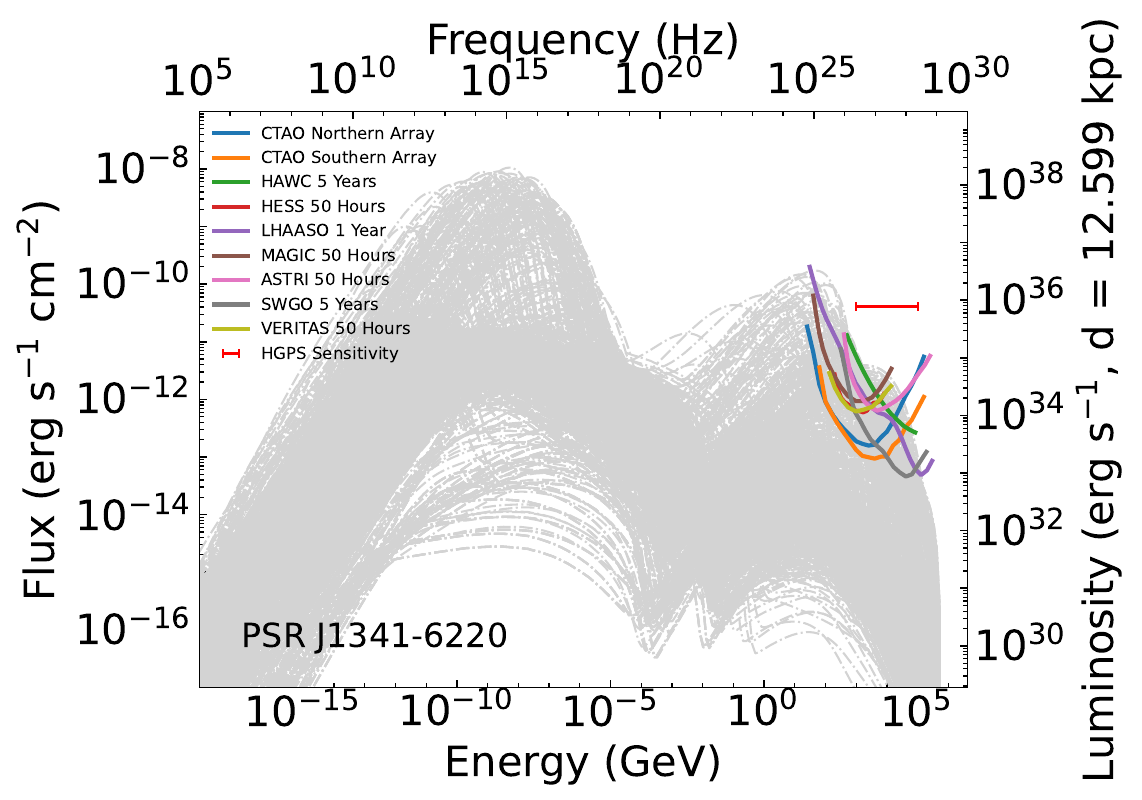}
   \caption{
Predicted set of SEDs for the PWNe of PSR J1341-6220. 
}
    \label{sed_J1341}
\end{figure}

The 1512 modeled SEDs for the putative PWN of PSR J1341-6220 are presented in Fig. \ref{sed_J1341}, and the relevant physical parameters involved are also shown in Table \ref{tab2}. This pulsar has the largest characteristic age among the four candidates, and from its SEDs, we observed that the PWN of PSR J1341-6220 might have already entered the reverberation phase in some model realizations. The TeV fluxes of the potential PWN in most models cannot be detected except by CTA. In addition, because of the location of this source, only S-CTA is suitable for a dedicated observation. We find that none of the 1512 model realizations would lead to a detectable TeV PWN in the HGPS, in agreement with the absence of detection. 

We classified all the models into three categories according to their visibility on different TeV telescopes:
    \begin{enumerate}
        \item {[< S-CTA]}: Lower than the sensitivity of the CTA Southern Array. 
        \item {[< H.E.S.S.]} \& [> S-CTA]: Between H.E.S.S. and the CTA Southern Array sensitivities. 
        \item {[> H.E.S.S.]}: Higher than the H.E.S.S. sensitivity. 
    \end{enumerate}
The number of models in different categories is listed in Table \ref{visibility}. One can see that in the majority of the models (close to 83\% of all those studied), the TeV emission from the PWN is undetectable. The TeV fluxes predicted for the remaining 17\% can be detected by S-CTA, and 6.55\% yield a TeV flux that could be discovered in a 50-hour dedicated observation by H.E.S.S.. 

\begin{table}
\setlength\tabcolsep{4pt}
        \centering
 \scriptsize
        \caption{Tera-electron volt visibility of the models for potential PWNe of selected pulsars that respect the X-ray upper limits when available. }
        \label{visibility}
 \begin{tabular}{llll} 
                \hline
        \hline
        PSR   & [< S-CTA] & [< H.E.S.S.] \& [>S-CTA] & [> H.E.S.S.]  \\
        \hline
        J1341-6220 (3 kpc)  &1132 (74.87\%)  &102 (6.75\%) &191 (12.63\%) \\ 
        J1341-6220 (6 kpc) &1206 (79.76\%)  &99 (6.55\%) &203 (13.43\%) \\
        J1341-6220 (12.6 kpc)  &1254 (82.94\%)  &159 (10.51\%) &99 (6.55\%) \\
        \hline
        J1838-0537 (2 kpc) &0   &16 (1.06\%) &73 (4.83\%) \\
        J1838-0537 (4 kpc) &0   &93 (6.15\%) &200 (13.23\%) \\ 
        J1838-0537 (6 kpc) &34 (2.25\%)  &229 (15.15\%) &202 (13.36\%) \\ 
        \hline
        J1844-0346 (2.4 kpc)  &0  &0 &0 \\ 
        J1844-0346 (4.3 kpc)  &0  &0 &0 \\ 
        J1844-0346 (10 kpc)  &8 (0.53\%)  &17 (1.12\%) &0 \\ 
\hline
        \end{tabular}
 \tablefoot{The sensitivities of S-CTA and H.E.S.S. utilized for comparison are $2.30\times10^{-13}$ and $1.94\times10^{-12}$ erg cm$^{-2}$ s$^{-1}$, respectively, which are directly integrated from the corresponding curves in the energy range of $1-10$ TeV.}
\end{table}

To understand which model parameters lead the separation in the different TeV visibility categories, we plotted the distributions of the nine parameters mentioned above and the radius $R_{pwn}$ and magnetic field $B_{pwn}$ for the 1000 random models under the detectable ([> S-CTA]) and undetectable ([< S-CTA]) categories. A Kolmogorov-Smirnov test (KS test) was also done to compare these two categories. Since the parameter values are randomly selected from their assumed ranges, their distributions are uniform. In contrast, a nonuniform distribution of a parameter in a particular category would show its influence over the PWNe TeV emission. 

\begin{figure*}
        \includegraphics[width=2\columnwidth]{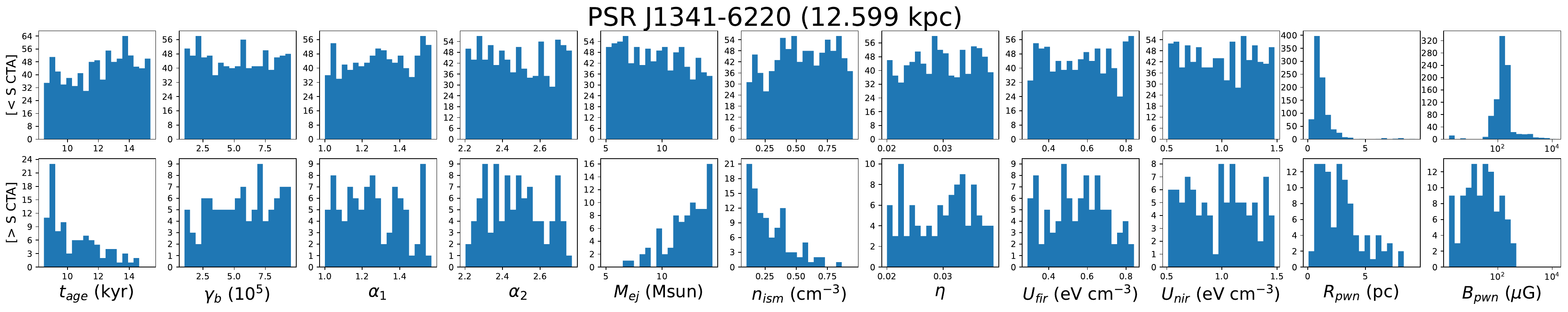}
   \caption{
Parameter distributions for the models analyzed for PSR J1341-6220 grouped into different categories. }
    \label{pars_distribution}
    \end{figure*}

As shown in Fig. \ref{pars_distribution} and Table \ref{KStest}, for the 897 models with a relatively low TeV emission ([< S-CTA]),  their $t_{age}$ and $n_{ism}$ are concentrated in the high end of the selected range, while $M_{ej}$ mostly appears in the lower end. The distributions of $R_{pwn}$ and $B_{pwn}$ are also different from models in the detectable category ([> S-CTA]), with more models having smaller radii and bigger magnetic fields. The distributions of the other parameters appear to be uniformly distributed, implying that their values have a secondary effect on the TeV radiation of this potential PWN. Most of the values of $R_{pwn}$ and $B_{pwn}$ are less than 4 pc and between 40 and 300 $\mu$G, respectively. Models with a high TeV emission ([>S-CTA] ) show the opposite of these distribution concentrations.

\begin{table*}
        \centering
 \scriptsize
        \caption{Kolmogorov-Smirnov test for the distributions of 11 parameters in different categories. }
        \label{KStest}
        \begin{tabular}{lccccccccccc} 
                \hline
        \hline
                    \textbf{Parameters} &$t_{age}$ &$\gamma_b$ &$\alpha_1$ &$\alpha_2$ &$M_{ej}$  &$n_{ism}$ &$\eta$ &$U_{fir}$ &$U_{nir}$ &$R_{pwn}$ &$B_{pwn}$\\
                \hline
  \multicolumn{3}{l}{\textbf{PSR J1341-6220}} & & & & & & & & &\\
            p-value Cat. 0 and Cat. 1: &$<10^{-10}$ &0.43 &0.06 &0.31 &$<10^{-10}$ &$<10^{-10}$ &0.54 &0.41 &0.90 &$<10^{-10}$ &$<10^{-10}$\\
                \hline
        \end{tabular}
 \tablefoot{In the table, Cat. 0 relates to random models with a flux [< S-CTA], and
 Cat. 1 relates to random models with a flux [> S-CTA],
and we refer to the p-value for a common parent population of the two quoted distributions. }
\end{table*}

Table \ref{KStest} shows that the results of the KS test for the distributions of each parameter support that the distributions of $t_{age}$, $n_{ism}$, $M_{ej}$, $R_{pwn}$, and $B_{pwn}$ for the different TeV visibility categories are not consistent with them being born from the same parent population, ruling out the null hypothesis with a p-value of smaller than $10^{-10}$. The [< S-CTA] category has older realizations than [>S-CTA], and if only free expansion is considered, then the PWN should have a larger radius and a weaker magnetic field. Figure \ref{pars_distribution} shows that it is the other way around, which means that the PWN has indeed entered a compression process in the framework of these models. Moreover, $M_{ej}$ and $n_{ism}$ will affect the age of the PWN from the free expansion phase to the reverberation phase, and the magnetic field grows enough to burn off energetic electrons when the PWN is compressed, which in turn affects the IC radiation and reduces the TeV radiation. All of this promotes further theoretical studies with models capable of coping better with the reverberation processes (see \citealt{bandiera2023reverberation}) as well as a dedicated observation using S-CTA, which will help further constrain our models and test the above conclusions. 

We also investigated scenarios with varying distance assumptions, similar to our approach for PSR J1208-6238. In cases of 3 and 6 kpc, nearly all models predict a PWN that remains undetectable in the HGPS. After excluding models with fluxes exceeding the HGPS sensitivity (only four models at 6 kpc and 87 models at 3 kpc), we obtained the number of models in the various categories specified in Table \ref{visibility}. Despite the larger flux with decreasing distance, the majority of the models indicate that this PWN remains undetectable by currently observing facilities (>74\%) even in the case that the uncertainty results in being in favor of increasing the flux on Earth. Thus, the level of distance uncertainty will have a minimal effect on the generic conclusions obtained.

\subsection{PSR J1838-0537}
\label{secJ1838}

PSR J1838-0537 is also a young and energetic radio-quiet $\gamma$-ray pulsar with a characteristic age of 4.89 kyr and a spin-down luminosity of $6.02 \times 10^{36}$ erg/s \citep{manchester2005australia}. The distance to this pulsar is also uncertain, and as in \cite{pletsch2012psr, albert2021evidence}, we chose 2.0 kpc. This value is based on the observed correlation between the $\gamma$-ray luminosity and $\dot{E}$ of a pulsar, and its uncertainty brings an obvious caveat, too. 

In the region where this pulsar is located, there is a TeV source, HESS J1841-055, that was also observed by HAWC as eHWC J1839-057 \citep{abeysekara2020multiple}. The potential PWNe of PSR J1838-0537 and PSR J1841-0456, as well as the supernova remnant (SNR) Kes 73, are suspected to be contributors to this VHE source \citep{gomez2020predicting}. The \cite{magic2020studying} tentatively investigated the physical nature and origin of the $\gamma$-ray emission from HESS J1841-055 and proposed leptonic and hadronic multi-source models. In their leptonic scenario, rather than IC radiation, bremsstrahlung is dominant, which is inconsistent with the usual results of PWN models. In their model, the TeV source is powered by one or several PWNe relics, so the IC emission efficiency is considered to be significantly higher than synchrotron emission, which can explain why there is no bright synchrotron nebula. 

\begin{figure}
\centering
    \includegraphics[width=.99\columnwidth]{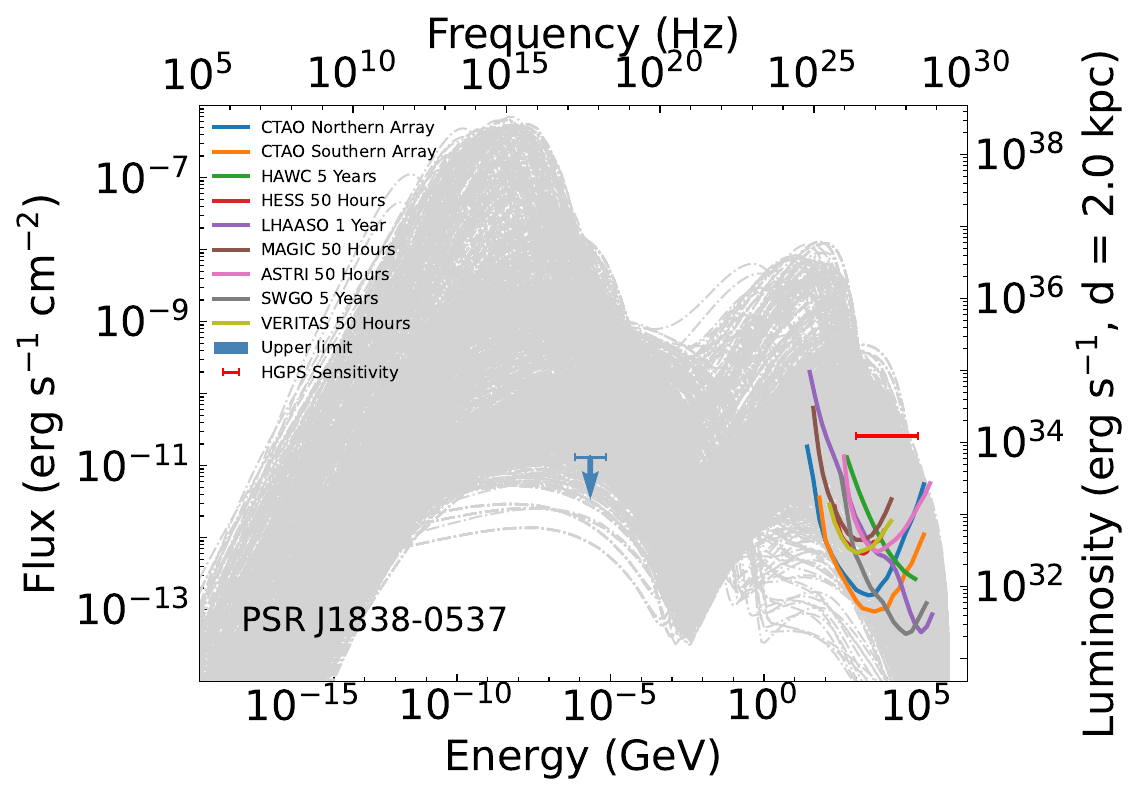}
   \caption{
Predicted set of SEDs for the PWN of PSR J1838-0537.
}
    \label{sed_J1838}
\end{figure}

However, PSR J1838-0537 is unlikely to leave such a PWN relic at such a young characteristic age. In addition, \cite{magic2020studying} considered the X-ray emission of the possible contributors (including nonpulsation X-ray flux from PSR J1838-0537) to HESS J1841-055 and used their total X-ray flux as the X-ray upper limit of this VHE extended source. We adopted this X-ray upper limit for the potential PWN of PSR J1838-0537 in this work as well. Possible SEDs and related physical parameters are shown in Table \ref{tab2} and Fig. \ref{sed_J1838}. 

The TeV radiation in the majority of the 1512 different models can be detected by H.E.S.S. (it might even have been detected already in HGPS as HESS J1841-055), and almost all of them are detectable with CTA. However, only in a small percentage of models does the X-ray emission respect the corresponding upper limit. In addition, in a small portion of models, this PWN has also entered the reverberation phase according to their SEDs. 

After excluding all the models that exceeded the X-ray upper limit, we divided the remaining 89 of them into the same three categories according to their detectability by different facilities, as described above. We found that most of them belong to [> H.E.S.S.], as displayed in Table \ref{visibility}. Only 17 of these 89 models respecting the X-ray upper limit come from the 1000 random model realizations, and all of them belong to the very luminous class [> H.E.S.S.], although none exceed the HGPS sensitivity. A dedicated deep observation of this source using CTA will be necessary in order to try to disentangle the different contributors in the region. 

Similar to our previous analyses, we also examined the impact of the distance uncertainty. For distances of 4 kpc and 6 kpc, many of the 1512 models studied (293 models for 4 kpc and 431 models for 6 kpc) comply with the X-ray upper limit. Table \ref{visibility} refers to their distributions in the various categories considered according to their detectability in the different observing facilities. Out of the 293 (431) models, none (only a small number) resulted in an undetectable PWN for S-CTA. However, there are still no models that can predict a detectable PWN in the HGPS as the assumed distance increases. In conclusion, the putative PWN powered by PSR J1838-0537 can likely be identified by S-CTA (and even possibly by H.E.S.S.) despite the considerable uncertainty regarding its distance. If the real distance exceeds 2 kpc, it is highly unlikely that HESS J1841-055 is the TeV counterpart of the putative PWN; instead, this PWN may at most be one of several contributors to this VHE source.

\subsection{PSR J1844-0346}
\label{secJ1844}

PSR J1844-0346 is also a $\gamma$-ray pulsar. It was found in the {\it Fermi}-LAT blind search survey and has a characteristic age of 11.6 kyr and a spin-down luminosity of 4.25$\times10^{36}$ erg s$^{-1}$ \citep{clark2017einstein}. Its distance is also unknown. 

HESS J1843-033, eHWC J1842-035, LHAASO J1843-0388, and TASG J1844-038 are VHE sources detected in this region. \cite{amenomori2022measurement} found that the energy spectrum of these VHE sources can be well fitted with a power-law function with an exponential cutoff. In the HGPS, \cite{abdalla2018hess} found that HESS J1843-033 consists of two merged offset components (HGPSC 83 and HGPSC 84), which seems to imply that this TeV source may have multiple origins. Because of their spatial proximity, PSR J1844-0346 and the radio SNR G28.6-0.1 were suspected of being the origin of the VHE emission. However, the association between PSR J1844-0346 and SNR G28.6-0.1 is unlikely. The estimated distance of SNR G28.6-0.1 is 6 – 8 kpc \citep{devin2021multiwavelength}, and for it, the resulting transverse velocity of the pulsar needed to reach from the center of the SNR to its current position would be over 1400 km/s, which is much larger than the typical value of a few hundred kilometers per second (e.g., \citealt{devin2021multiwavelength}). The same transverse velocity argument was made for a more likely association of PSR J1844-0346 with the star-forming region N49 at 5.1 kpc. Alternative empirical estimates of the distance of PSR J1844-0346 are 2.4 kpc \citep{wu2018einstein}, obtained by assuming that the $\gamma$-ray luminosity scales as $\sqrt{\dot{E}}$, and 
4.3 kpc, based on the empirical relation obtained for $\gamma$-ray pulsars \citep{parkinson2010eight}. None of these are certain, however, and they are herein considered to span the possible results.

\cite{zyuzin2018x} used nearly 100 ks exposure data from the Swift X-Ray Telescope (XRT) to discover an X-ray counterpart candidate of this $\gamma$-ray pulsar. The candidate has an estimated unabsorbed flux of $2.2^{+1.3}_{-0.4}\times10^{-13}$ erg cm$^{-2}$ s$^{-1}$ (0.3 - 10 keV). \cite{devin2021multiwavelength} searched for radio or X-ray counterparts around PSR J1844-0346 that could indicate a possible PWN but did not find any, and they do not provide any diffuse upper limit to compare our results with. We took the X-ray flux of \cite{zyuzin2018x} as an X-ray upper limit for this pulsar's potential PWN in case the latter can be regarded as point-like. Although if the latter is more extended and diluted, which is not expected at this age, the flux can be larger.

\begin{figure}
\centering
    \includegraphics[width=0.99\columnwidth]{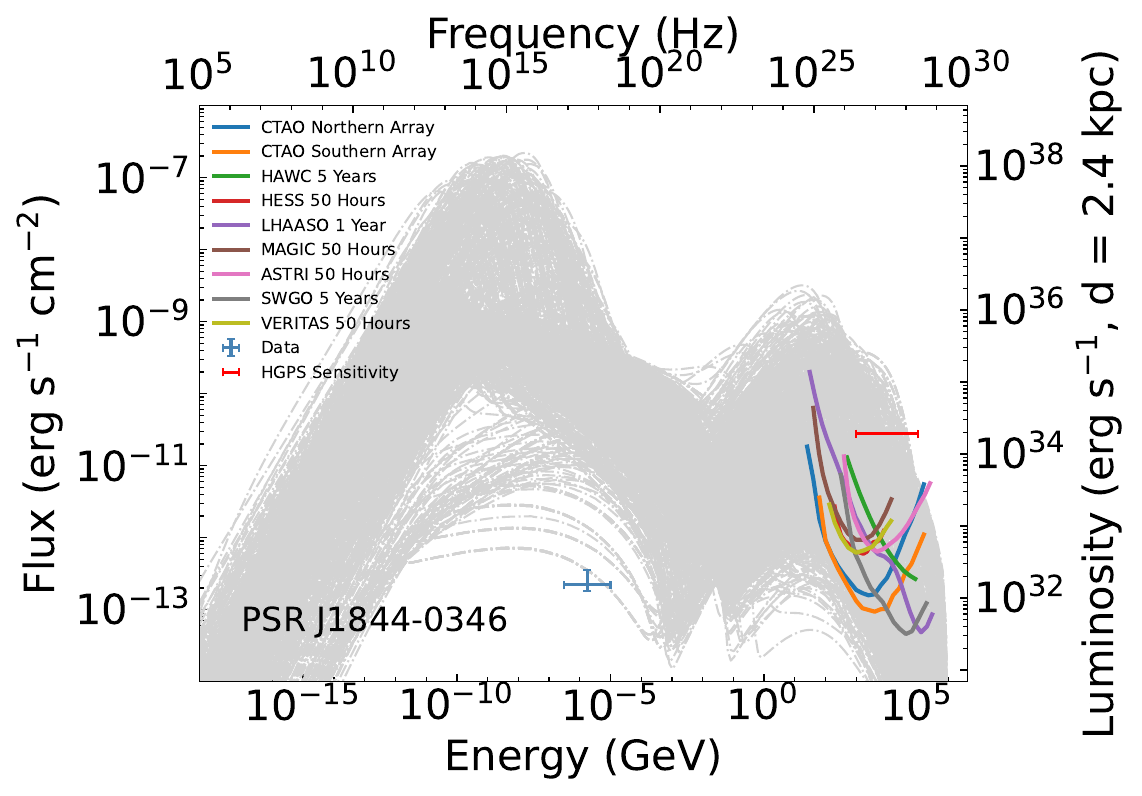}
    \includegraphics[width=0.99\columnwidth]{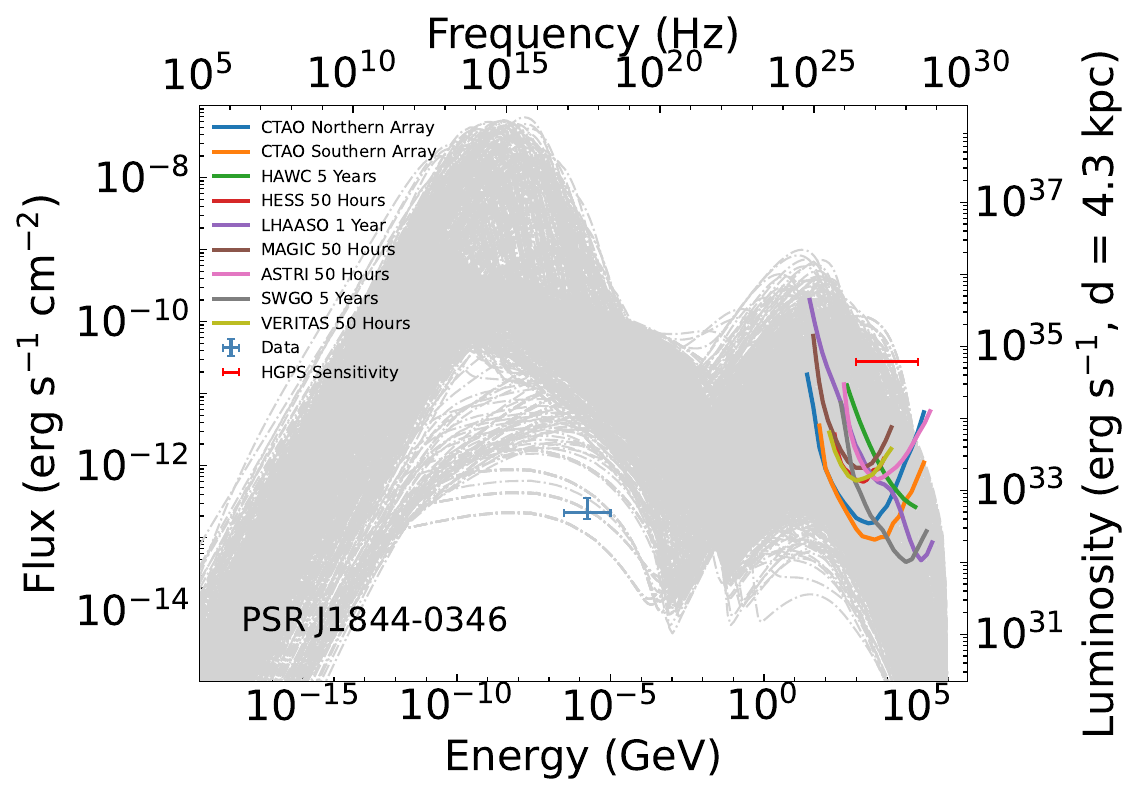}
    \includegraphics[width=0.99\columnwidth]{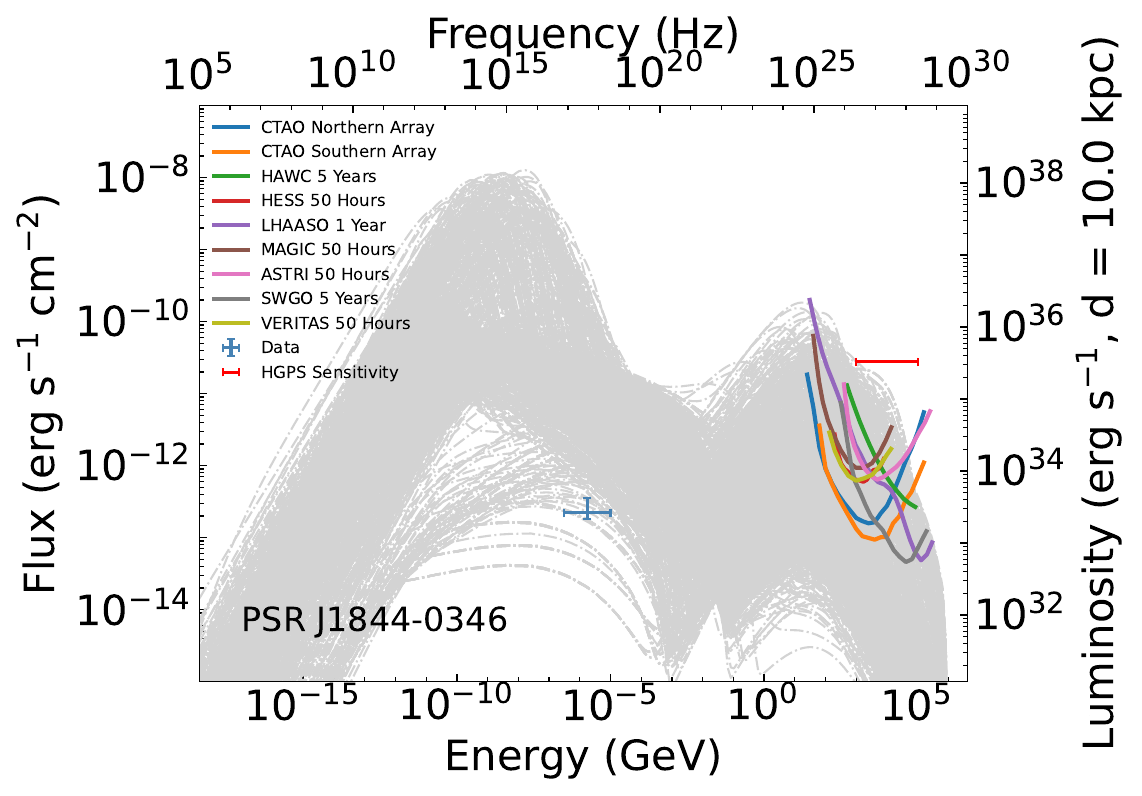}
   \caption{
Predicted set of SEDs for the PWN of PSR J1844-0346 located at 2.4, 4.3, and 10 kpc, respectively. 
}
    \label{sed_J1844}
\end{figure}

Similar to the other three candidates, the 1512 different model realizations with distances assumed to be 2.4 kpc can also be seen in Table \ref{tab2} and in the top panel of Fig. \ref{sed_J1844}. The characteristic age of 11.6 kyr allowed us to see that many of the possible PWNe have entered the reverberation phase. Due to the closer distance and the more energetic $\dot{E}$, the overall energy flux of PSR J1844-0346's PWN is higher compared to PSR J1341-6220's PWN, and this leads to the TeV emission being detectable by CTA in the most (> 85\%) models and by H.E.S.S. in close to 30\%. However, if we are to assume the pulsar's X-ray flux as an upper limit for the PWN (which would only make sense for small PWN), none of the SEDs predicted by our models lead to detectable TeV emission except with an assumed distance of $\gtrsim 10$ kpc (see Fig. \ref{sed_J1844} and Table \ref{visibility}).

Ultimately, this PWN would not be an obvious candidate for S-CTA or H.E.S.S.. However, this conclusion is weakened when the true distance of this PWN is much larger than 2.4 kpc or its X-ray diffuse emission is much larger than the X-ray flux of the pulsar that we adopt here.

\section{Conclusions}
\label{conclusion}

In this work, we aimed to discover the detectable TeV PWNe. With this in mind, we selected four pulsars and studied them as candidates to produce a detectable TeV PWN based on their locations on the pulsar tree, the MST of the pulsar population, as we have noted that the majority of the known TeV PWNe are closely grouped together.

Using the observational results of the four candidates and assuming within reasonable ranges some model parameters when unknown, we used our leptonic model to predict a large set of SEDs for the putative PWNe of each of these four pulsars. We then proceeded to analyze the set, and by comparing the TeV fluxes predicted by the models that respect the X-ray observations when existing with the sensitivities of current and future instruments, we concluded how likely it is for each of the PWNe to be detectable. 

We find that the TeV emission of the potential PWNe of PSR J1208-6238 and PSR J1341-6220 is expected to be relatively low, which is consistent with the fact that we have not detected their TeV counterparts yet. This is also the case if the pulsars are located at larger distances than assumed, which can happen within the associated uncertainty of this parameter. The TeV emission predicted by the models for PSR J1208-6238 at different distances is even lower than the sensitivity of S-CTA. The other two pulsars, PSR J1838-0537 and PSR J1844-0346, have a relatively higher TeV emission from their putative PWNe, and our results tend to suggest that they could be detectable by S-CTA or even by H.E.S.S. in sufficiently long dedicated observations. However, only a reduced number of model instances could lead to detectable TeV emission without violating X-ray constraints if they are used as PWN upper limits.

The above results also allowed us to understand the convenience and caveats of the pulsar tree in pulsar population analysis and pinpointing detectable PWNe. The pulsar tree that we used to select candidates only contains intrinsic information about the pulsars themselves, with no reference to the environment or the distance to them. Even with just this kind of information, the grouping of all PWNe is impressive (see Fig. \ref{fig1}). However, we find that the location of a yet nondetected (and in principle likely undetectable) PWN, among others that are observed, does not necessarily imply a different behavior for the system. Finally, we note that most of the nondetected PWNe pertain to the same branch within the region of the pulsar tree where most of the PWNe are located, which is formed by the less energetic pulsars of the set. 

Given that we had a thousand models for each PWN, we could also test the general correlations without imposing any constraints. We used Pearson’s and Spearman’s tests to analyze whether $L_{TeV}$ is correlated with any of the model parameters. The largest correlation coefficients were obtained in the case of J1208-6238 for the pairs ($\alpha_2$, $L_{TeV}$), with $r=-0.61$ (Pearson) and $r=-0.69$ (Spearman). This represents a moderate to strong correlation. All other model parameters present milder correlation coefficients (typically well below $0.3$). We encountered a number of factors when searching for larger correlation coefficients. On the one hand, the cross-influence of all parameters is a well-known effect. For instance, if we keep the electron spectrum fixed, the higher the energy density is, the higher the TeV luminosity will be. However, if at the same time the magnetization is increased (so that the synchrotron losses are higher, affecting the electron population) or the age is increased, the electrons are more or less cooled as a result, and even for a higher photon density, there could be a lower TeV luminosity. As we have several parameters and all vary at once, correlations are correspondingly less clear. This is a widely known effect. On the other hand, the appearance of models entering into reverberation also affects correlations, as this phase gives rise to new phenomenology when the medium compresses the PWN shell. Our models have also been tested by comparing the parameters of those models that generate a large $L_{TeV}$ (in regard to the observational sensitivities) with those models that do not. We did this by testing via a KS test the null hypothesis that states that the distribution of parameters of both TeV luminous and dim PWNe are consistent with having the same parent population. The case of PSR J1341-6220 can be considered an example of this investigation. As for the distribution of $M_{ej}$, most models predicting a high TeV emission have a relatively larger $M_{ej}$ (and a smaller $n_{ism}$), which means a larger Sedov time ($t_{Sed} \propto M_{ej}^{5/6}\cdot n_{ism}^{-1/3}$). This is not obvious in the Pearson's and Spearman's tests (which in this case have a Pearson coefficient of 0.24 and a Spearman coefficient of 0.11), but it is captured by the KS test (see, e.g., Fig. 4 and Table 4).

In conclusion, this study provides a case study of pulsar population analysis using the pulsar tree, as well as two promising PWNe that could be detected with CTA and even H.E.S.S.. Therefore, they are also worthy of further observation in the TeV and other energy bands. Furthermore, we examined the impact of varying intrinsic pulsar parameters on the TeV radiation of their young PWNe. This work underscores the potential for advancing our understanding of pulsar characteristics and PWN evolution through high-energy observations.

\begin{acknowledgements}
This work was supported by the National Scholarship Council (PhD fellowship from the China Scholarship Council (CSC) (No. 202107030003))
and by the grant PID2021-124581OB-I00 of MCIU/AEI/10.13039/501100011033 and 2021SGR00426. 
This work was also supported by the program Unidad de Excelencia María de Maeztu CEX2020-001058-M and also supported by MCIU with funding from European Union NextGeneration EU (PRTR-C17.I1). 
We thank an anonymous referee for  helping us improve the manuscript.
\end{acknowledgements}

%
   \bibliographystyle{aa} 
   \bibliography{reference} 
%

\end{document}